\documentclass[letterpaper]{jpconf}

\usepackage{graphicx}
\usepackage[square,sort&compress]{natbib}
\begin{document}
\title{E/B mode mixing}
\author{Emory F Bunn}
\address{Physics Department, University of Richmond, Richmond, VA  23173}
\ead{ebunn@richmond.edu}
\begin{abstract}
In future microwave background polarization experiments, particularly
those that aim to characterize the $B$ component, careful attention
will have to be paid to the mixing of $E$ and $B$ components due
to finite sky coverage and pixelization.  Any polarization map
can be decomposed into ``pure $E$,'' ``pure $B$,'' and ``ambiguous''
components.  In practice, since the $B$ component is expected to
be much weaker than the $E$ component, nearly all of the recoverable $B$
information is contained in  the pure $B$ component.  The amount
of $B$ information lost to ambiguous modes can be estimated
in simple ways from the survey geometry and pixelization.  Separation
of any given map into pure and ambiguous components can be done
by finding a basis of pure and ambiguous modes, but it is often
more efficient to ``purify'' the map directly in real space by solving
a certain differential equation to find the ambiguous component.
This method may be useful in conjunction with power spectrum estimation
techniques such as the pseudo-$C_l$ method.  
\end{abstract}

\section{Introduction}
Future cosmic microwave background (CMB) polarization experiments will
have as a key goal the detection and characterization of the
pseudoscalar $B$ polarization component \cite{zalsel,selzal,kkslett,kks}.  
This component is predicted
to be weaker than the scalar $E$ component by an order of magnitude or more
over all angular scales.  
In any claimed detection of the $B$ component,
therefore, it is essential that contamination from $E$ modes be strictly
limited.

The separation of a spin-two
polarization map into $E$ and $B$ components is closely
analogous to the more familiar separation of a vector field into curl-free
and divergence-free components.  In both cases, the separation can be done
perfectly, with no ``leakage'' of one component into the other, as long
as the data set covers a region with no boundary (i.e., the entire sky).
In the more realistic case where some portion of the sky is unobserved,
there is ambiguity in the EB separation \cite{LCT,bunn,bunnetal}.

Given a polarization map, the primary science goal is to extract the 
power spectra $C_l^E,C_l^B$ 
for the $E$ and $B$ components.  In principle, this goal
can be achieved without
performing an actual separation of the map into components: we
can simply compute the likelihood
function $L(C_l^E,C_l^B)$ for the data set and estimate the power spectra
from it.  For a Gaussian theory, the likelihood function encodes all of
the information in the data.  This procedure would therefore naturally
account for any ambiguity arising from imperfect $E$/$B$ separation and
give the best possible error bars on the recovered power spectra.
In practice, however, actual separation of a map into $E$ and $B$ components
will surely be important for a variety of reasons, including
tests for foreground contamination, non-Gaussianity,  and systematic
errors.  Last but not least, in any experiment that claims to detect $B$ modes,
actually displaying the $B$ mode map will be an essential part of 
convincing the community of the reality and importance of the detection.
(For example, consider the original COBE detection: although the
key science was contained in the two-point correlation function and power
spectrum estimates, the actual real-space maps were invaluable in 
convincing the world of the validity and importance of the results.)

Consideration of issues related to $E$/$B$ separation is important in
experiment design and optimization as well.  For example, the ambiguity
in $E$/$B$ separation significantly alters the optimal
tradeoff between sky coverage and noise per pixel in a degree-scale $B$ mode
experiment \cite{bunn}.

\section{Pure and ambiguous modes}

\begin{figure}[t]
\centerline{
\raisebox{0.4in}{(a)} \includegraphics[width=3in]{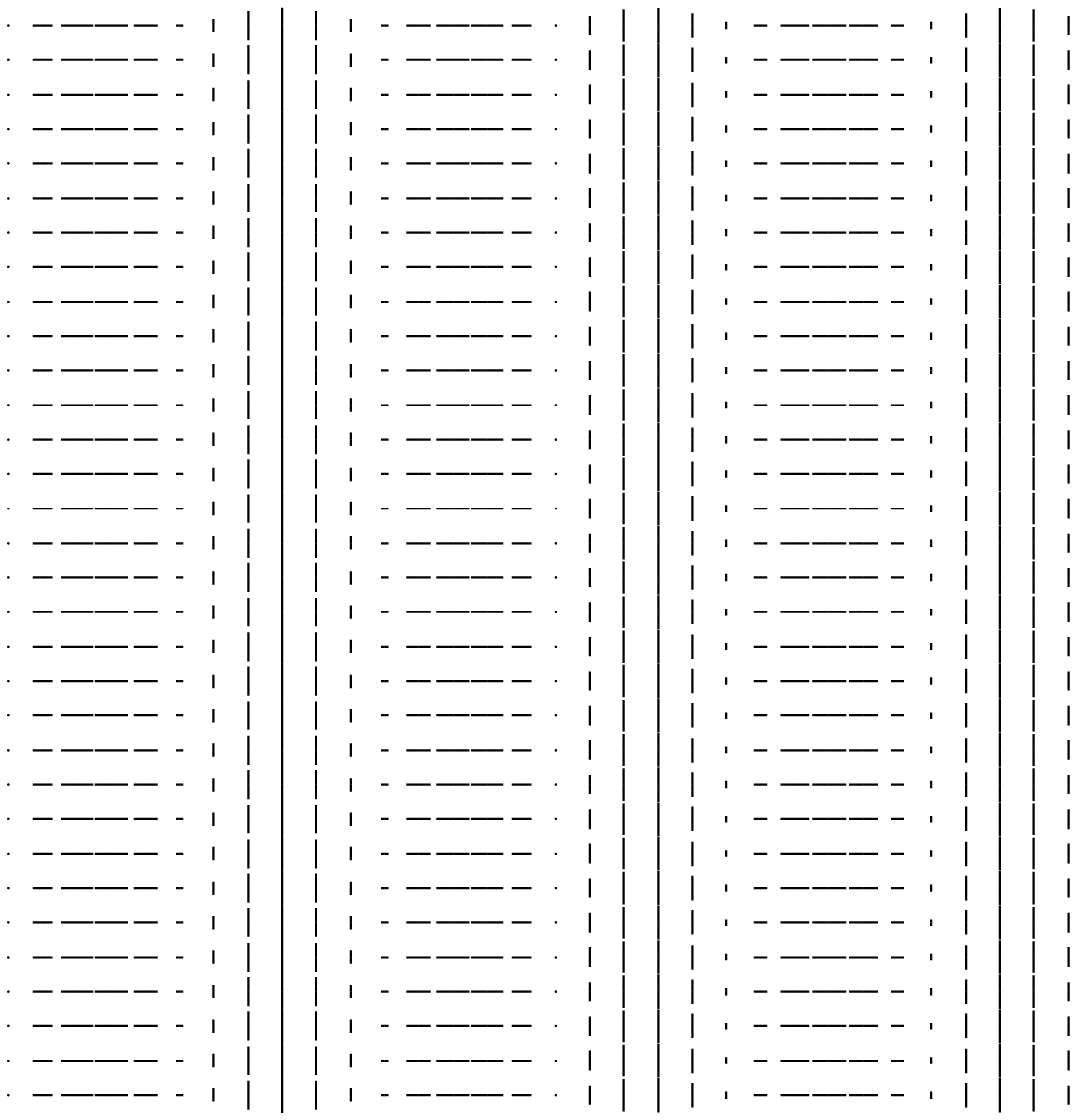}\quad
\raisebox{0.4in}{(b)} \includegraphics[width=3in]{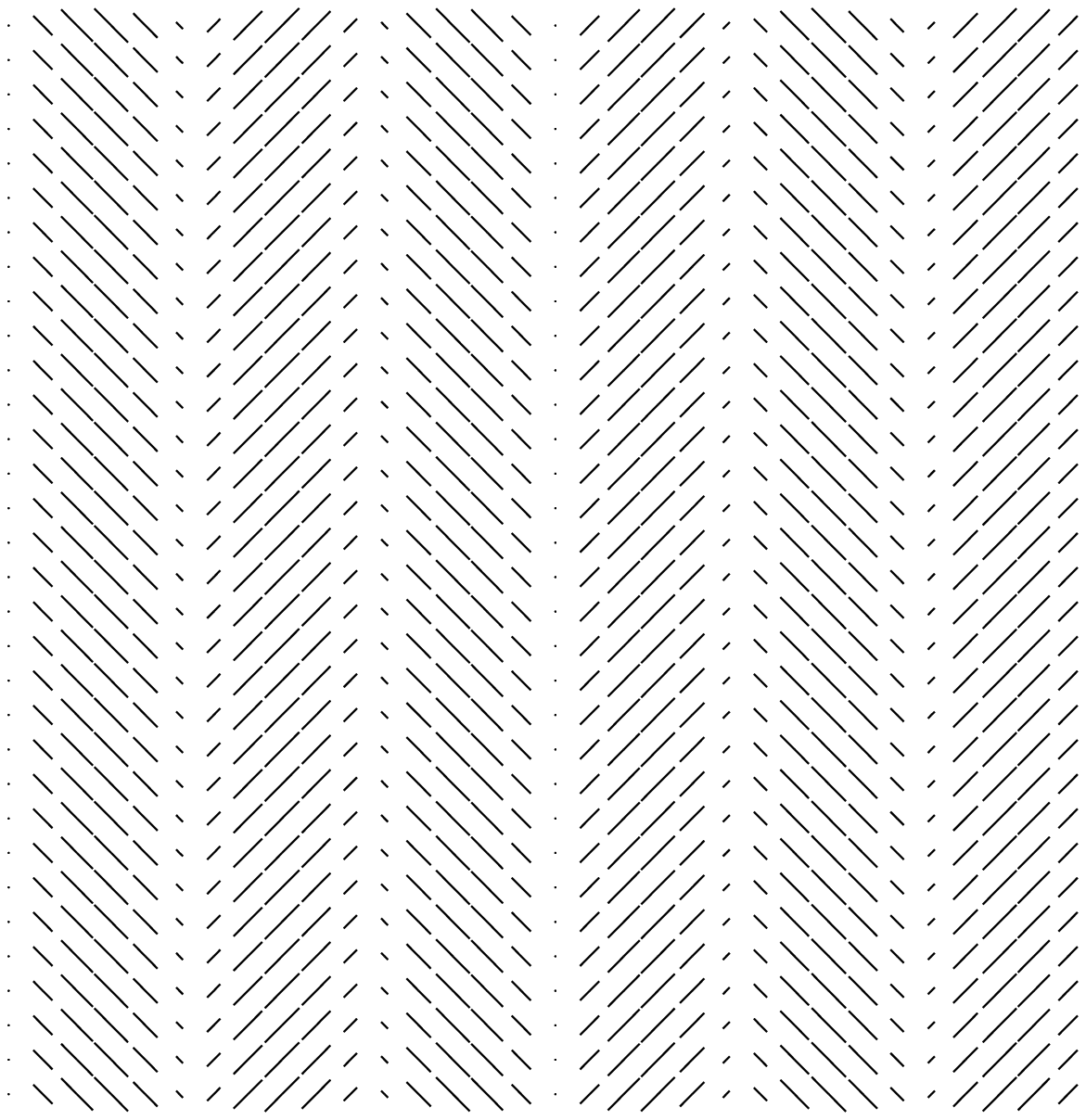}
}
\caption{A pure $E$ Fourier mode (a),
and a pure $B$ mode (b).}
\label{fig:ebfourier}
\end{figure}

The $E$/$B$ decomposition is easiest to understand in Fourier space.  For
any given wavevector ${\bf k}$, define a coordinate system $(x,y)$ with
the $x$ axis parallel to ${\bf k}$, and compute the Stokes parameters
$Q,U$.  An $E$ mode contains only $Q$, while a $B$ mode contains only $U$.
In other words, in an $E$ mode, the polarization direction is
always parallel or perpendicular to the wavevector, while in a $B$ mode
it always makes a $45^\circ$ angle, as shown in Figure \ref{fig:ebfourier}.

In a map that covers a finite portion of the sky, of course, the Fourier
transform cannot be determined with infinite $k$-space 
resolution.  According to 
the Heisenberg uncertainty principle, if the observed region
has size $L$, an estimate of an individual
Fourier mode with wavevector ${\bf q}$ will be a weighted
average of true Fourier modes ${\bf k}$ in a region around ${\bf q}$ of 
width $|{\bf k}-{\bf q}|\sim L^{-1}$.  These Fourier modes will all point
in slightly different directions, spanning a range of angles $\sim qL$.
Since the mapping between $(Q,U)$ and $(E,B)$ depends on the angle of the
wavevector, we expect the amount of $E$/$B$ mixing to be of order $qL$.  
In particular, this means that the largest scales probed by a given
experiment will always have nearly complete $E$/$B$ mixing.
This is unfortunate, since the largest modes probed are generally the 
ones with highest signal-to-noise ratio.  Typically, the noise variance
is about the same in all Fourier modes detected by a given experiment,
while the signal variance scales as $C_l$, which decreases as a function
of wavenumber.  (Remember, even a ``flat'' power spectrum
is one with $l^2C_l\sim$ constant.)

One way to quantify the amount of information lost in a given experimental
setup is to decompose the observed map into a set of orthogonal
modes consisting of {\it pure $E$} modes, {\it pure $B$} modes, and 
{\it ambiguous} modes \cite{bunnetal}.  A pure $E$ mode is orthogonal
to all $B$ modes, which means that any power detected in such a mode
is guaranteed to come from the $E$ power spectrum.  Similarly, pure $B$ modes
are guaranteed to come from the $B$ spectrum.  Any power 
detected in an ambiguous mode could have come from either $E$ or $B$.  In
practice, these modes are likely to be mostly $E$, and any $B$ contribution
they contain will be impossible to separate.  

\begin{figure}[b]
\centerline{\includegraphics[width=3in]{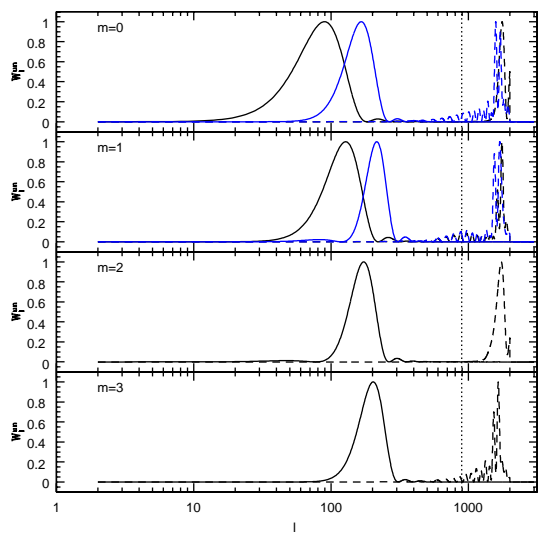}}
\caption{Window functions for $E$ and $B$ modes for a hypothetical experiment
in which measurements are made on a spherical cap.  The dashed lines
show the leakage of $E$ into $B$ or vice versa.  The leakage is caused
by pixelization and appears at the Nyquist scale, indicated by the vertical
dotted line.  For further details,
see ref.~\cite{bunnetal}, from which this figure is taken.}
\label{fig:pixel}
\end{figure}

In the limit of a map with infinitely dense pixelization, the $E$/$B$/ambiguous
decomposition depends only on the geometry of the observed region.
Orthonormal bases for the pure and ambiguous subspaces can be found
by finding eigenfunctions of a certain differential operator \cite{bunnetal}.
As expected from the heuristic argument above, the amount of information
lost to ambiguous modes is largest on the largest observable scales.  To be
specific, the ratio $N_{\rm amb}/N_{\rm pure}$ of ambiguous to pure modes
is $\sim 1$ on the largest observed scales and decreases as $1/k$
with increasing wavenumber $k$.

Pixelization also causes mixing of $E$ and $B$ modes.  We can understand
this intuitively by imagining a survey that observes a square 
region of sky.  Pixelization introduces a Nyquist wavenumber $k_{\rm Ny}$,
such that all modes with wavevector components greater than $k_{\rm Ny}$ are
aliased to modes with wavenumbers less than the Nyquist value. This
aliasing completely shuffles the directions of the wavevectors, so it
leads to essentially complete $E$/$B$ mixing.  In particular, consider an $E$ mode
with wavenumber close to the Nyquist value.  The mean-square amplitude
of this mode will be $C_{k_{\rm NY}}^E e^{-\sigma_b^2k_{\rm Ny}^2}$, where
$\sigma_b$ is the Gaussian beam width.  Due to aliasing, a significant
fraction of the power in this mode will appear to be part of a $B$ mode
with mean-square amplitude $C_k^B$ for some small wavenumber $k$.  
To avoid significant degradation of $B$ mode measurements, therefore, we should
pixelize on a fine enough scale (that is, make $k_{\rm Ny}$ large enough)
to satisfy the inequality
\begin{equation}
e^{-\sigma_b^2k_{\rm Ny}^2}<{C_k^B\over C_{k_{\rm Ny}}^E}.
\end{equation}
This typically implies oversampling the beam of the experiment
by a factor of $\sim 3$-$4$.

Figure \ref{fig:pixel} shows an example of the effects of pixelization
on the window functions for supposedly pure $E$ and $B$ modes.  The
dashed curves show the ``leakage'' window functions, that is, the contribution
of $E$ to a supposed $B$ mode and vice versa.  The leakage appears suddenly
at a particular scale, which is the Nyquist scale for the experiment
in question.  For further details, see ref.~\cite{bunnetal}.

\section{Efficient real-space decomposition}
Finding a basis of orthonormal pure and ambiguous modes is in principle
straightforward, but for a map with a large number of pixels and/or a 
complicated geometry, it can be computationally expensive.  In this section,
I sketch a method for decomposing a map into  pure $E$, pure $B$, and
ambiguous maps without finding such a basis.  The procedure consists
of two steps:
\begin{enumerate}
\item[1.] Do a na\"\i ve separation of the map into $E$ and $B$ components,
without worrying about the presence of ambiguous modes.  
At the end of this step, the original map will be written as the
sum of two terms, one of which consists of pure $E$ and ambiguous
modes, and one of which consists of pure $B$ and ambiguous modes.
\item[2.] ``Purify'' each of the two maps found in step 1.
\end{enumerate}
Step 1 can be done in a wide variety of ways.  In the flat-sky approximation,
one way is to pad the map out to a rectangular shape (using zeroes or
any other values for $Q,U$ in the padded regions).  One then performs 
a fast Fourier transform on the padded map, performs the $E$/$B$ decomposition
trivially in Fourier space, and transforms back. 
The resulting $E$ map will consist entirely of $E$ modes, although
not necessarily of pure $E$ modes -- that is, it will be a combination
of $E$ and ambiguous modes.  
If the flat-sky approximation is not appropriate, the same 
can be done in spherical harmonic space, e.g., with HEALPix \cite{healpix}.

We must now ``purify'' each of these two maps.  In an earlier
work \cite{bunnmn}, I described one way to do this, based
on Green functions; however, that method is not very efficient in practice.
I will now sketch a simpler and more efficient method.

Let us recall some
mathematical properties of the ambiguous modes \cite{LCT,bunnetal}.  
Any $E$ mode 
(pure or not) can be written in terms of a scalar potential $\psi$:
\begin{equation}
{\bf P}={\bf D}_E\psi.
\end{equation}
Here ${\bf P}$ represents the spin-two field $(Q,U)$, and ${\bf D}_E$
is a certain second-order differential operator.  Furthermore, all
$B$ modes satisfy the equation
\begin{equation}
{\bf D}_E^\dag\cdot{\bf P}=0.
\end{equation}
Since an ambiguous mode is both an $E$ mode and a $B$ mode, it
can be derived from a potential $\psi$ that satisfies
the differential equation ${\bf D}_E^\dag\cdot{\bf D}_E\psi=0$, which
is equivalent to
\begin{equation}
\nabla^2(\nabla^2+2)\psi=0.
\label{eq:biharmonic}
\end{equation}
Solutions to this equation are uniquely determined by 
the values of $\psi$ and its derivatives on the boundary of the
region, which can be expressed in terms of the data contained
in the $Q,U$ maps.  Furthermore, equation (\ref{eq:biharmonic})
can be solved numerically quite efficiently, for example by relaxation
methods.  

\begin{figure}[h]
\centerline{
\raisebox{0.3in}{(a)} \includegraphics[width=2in]{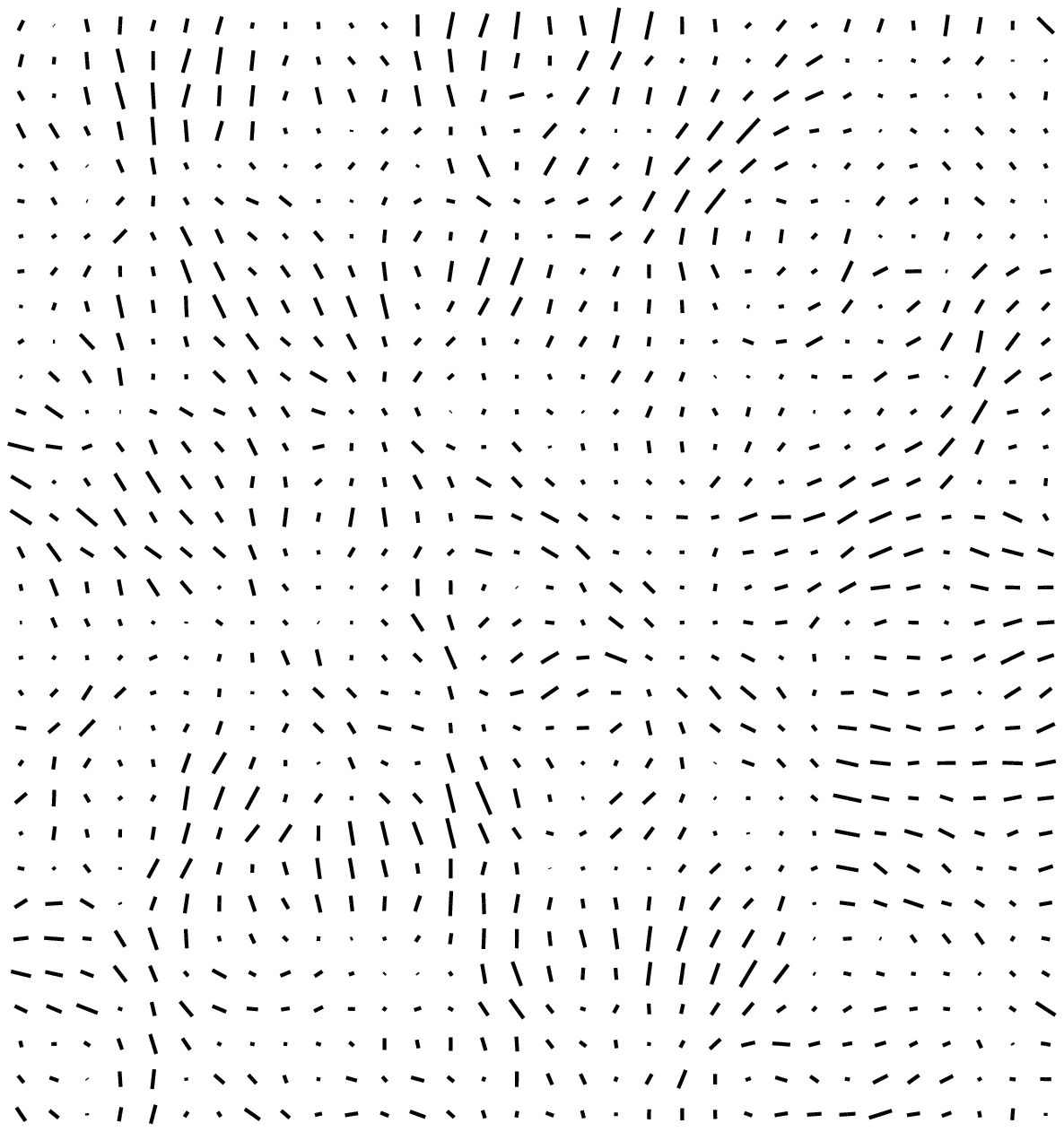}
\raisebox{0.3in}{(b)} \includegraphics[width=2in]{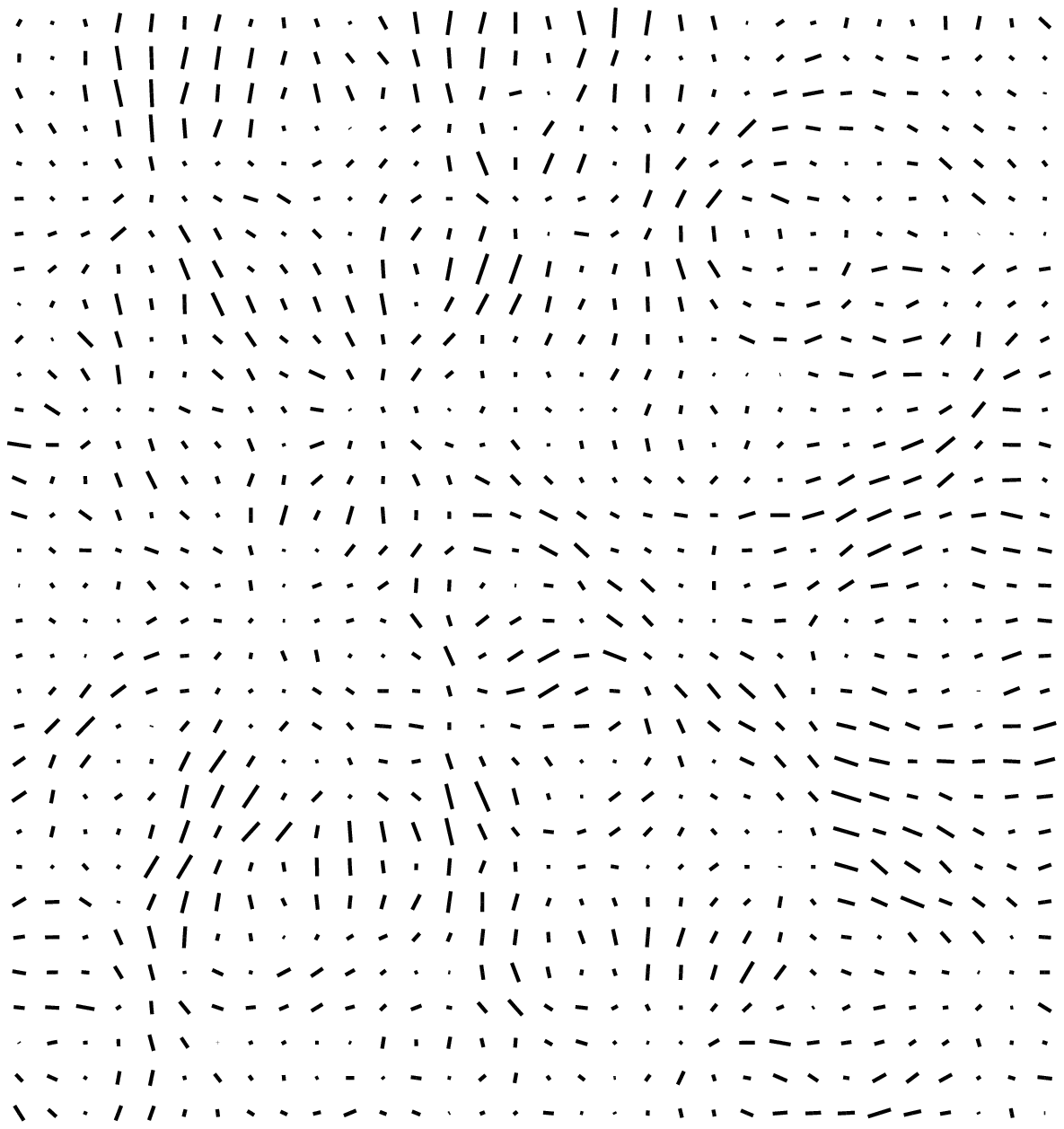}
\raisebox{0.3in}{(c)} \includegraphics[width=2in]{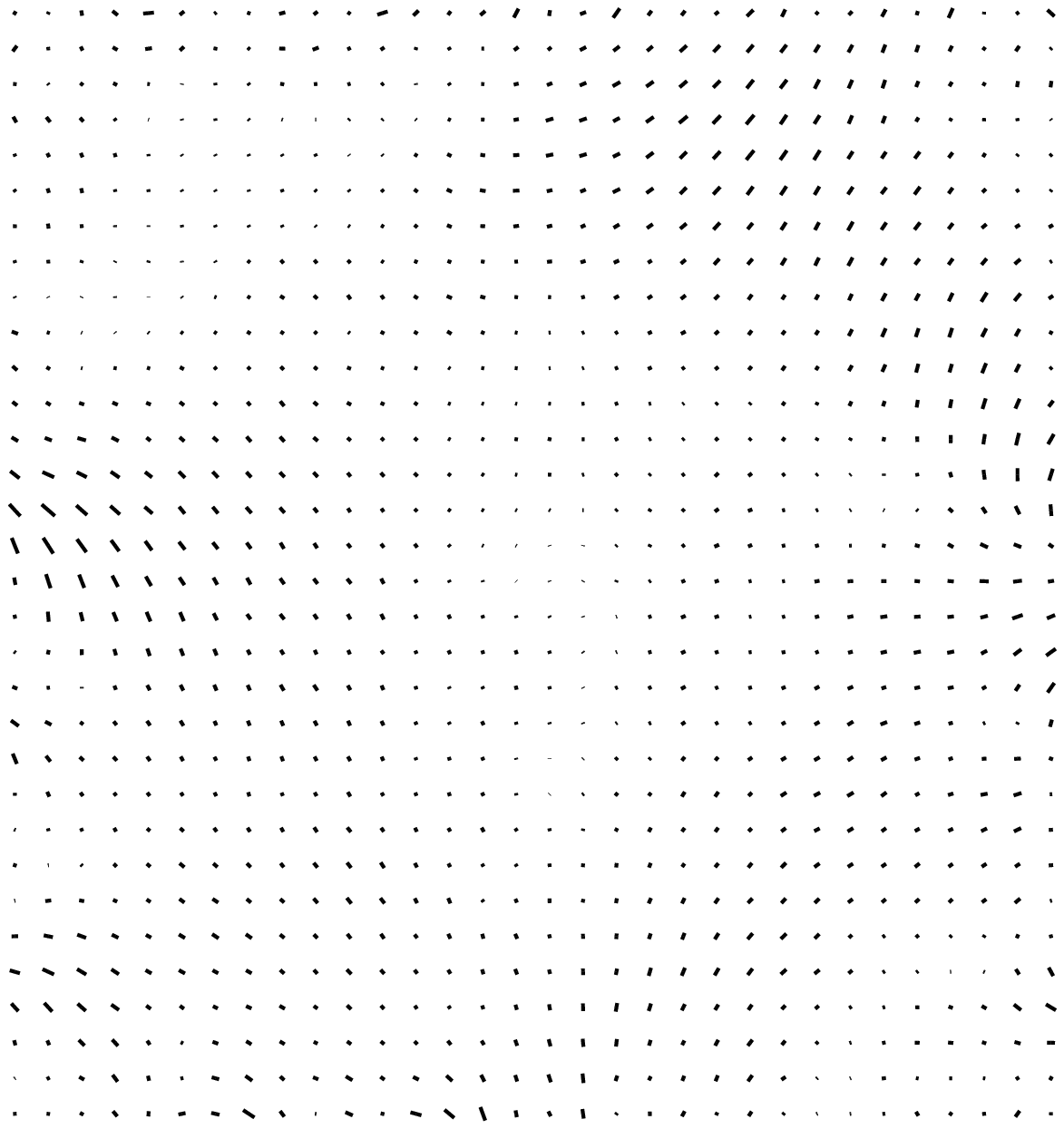}}
\centerline{
\raisebox{0.3in}{(d)} \includegraphics[width=2in]{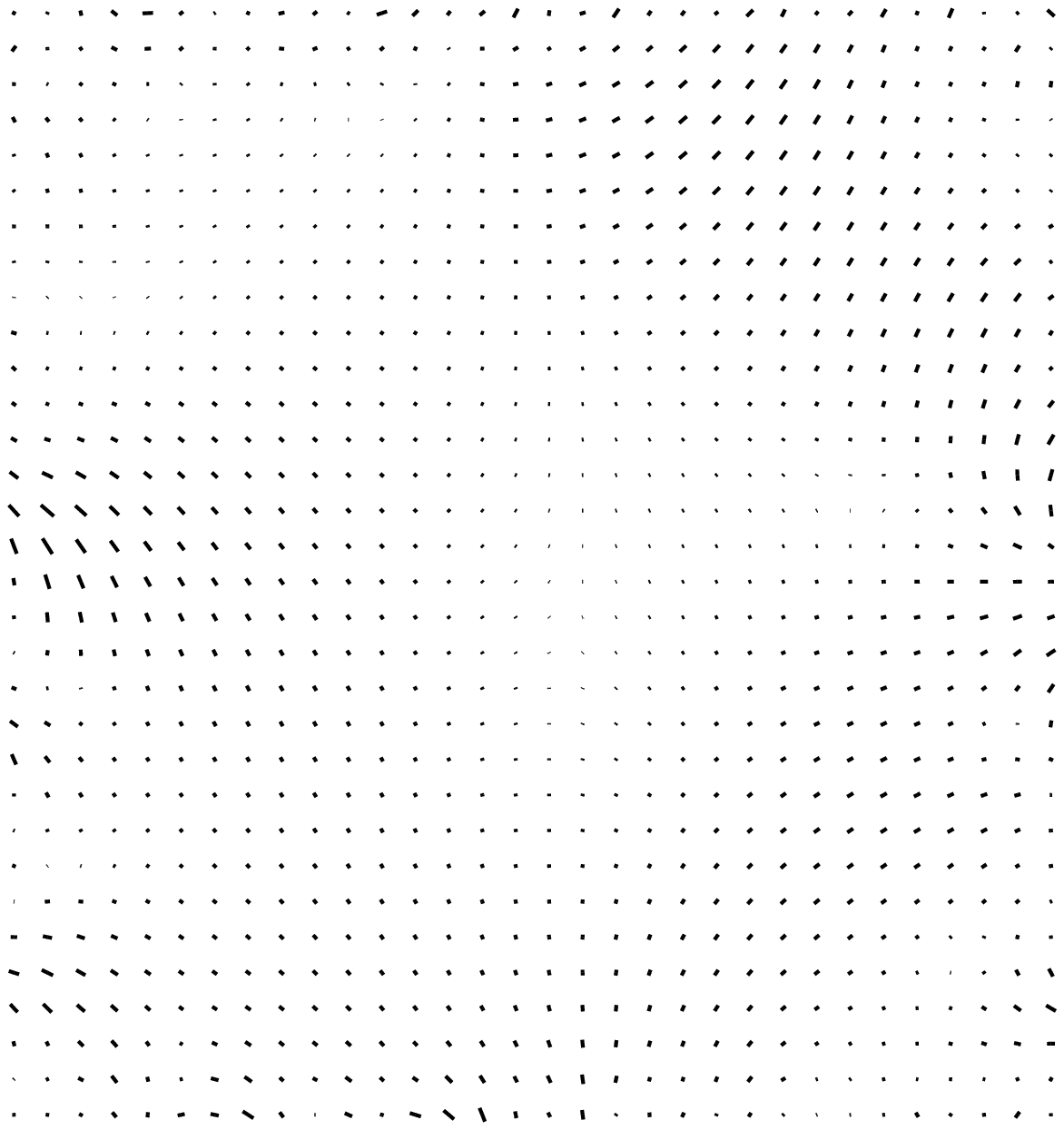}
\raisebox{0.3in}{(e)} \includegraphics[width=2in]{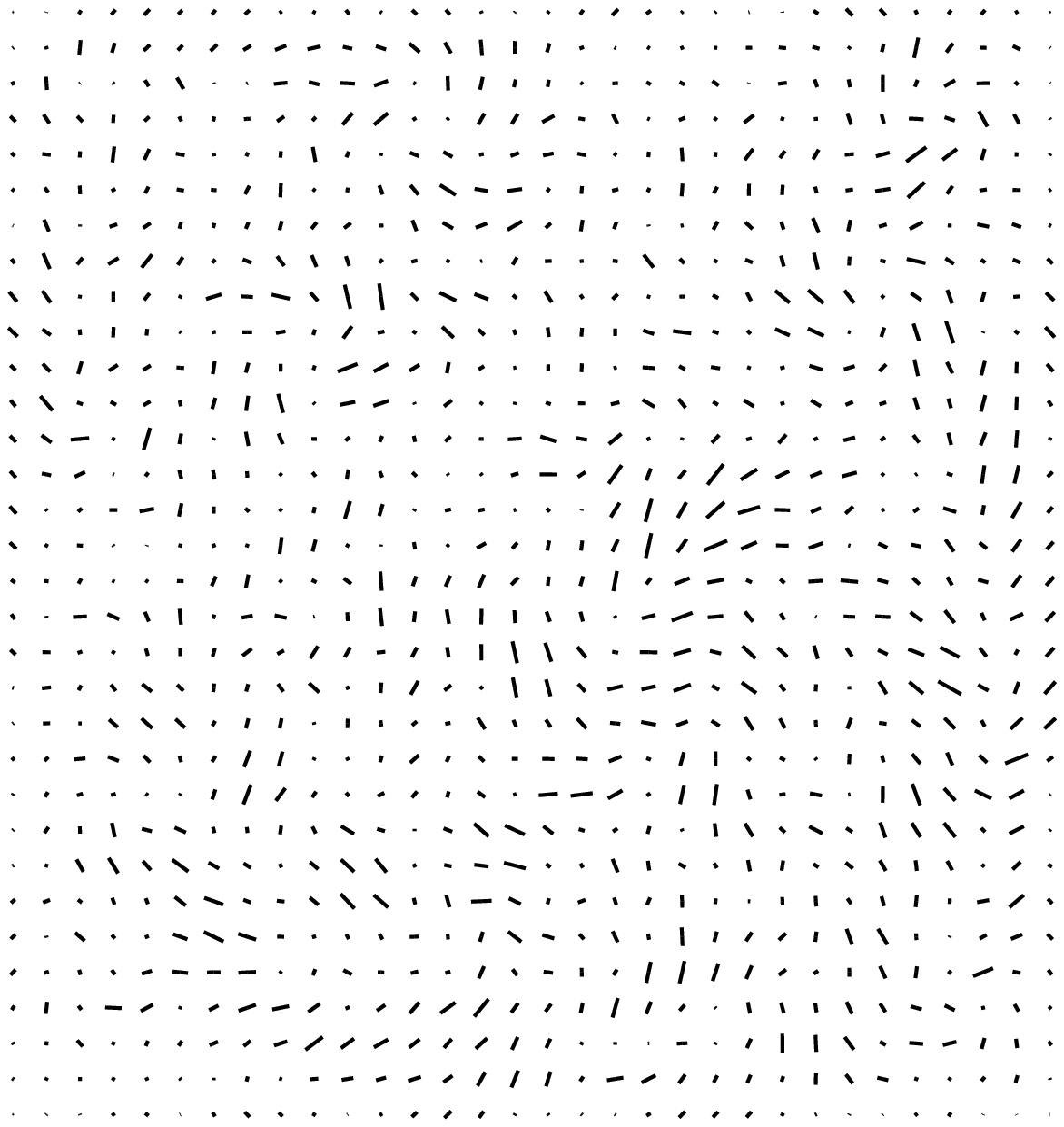}
\raisebox{0.3in}{(f)} \includegraphics[width=2in]{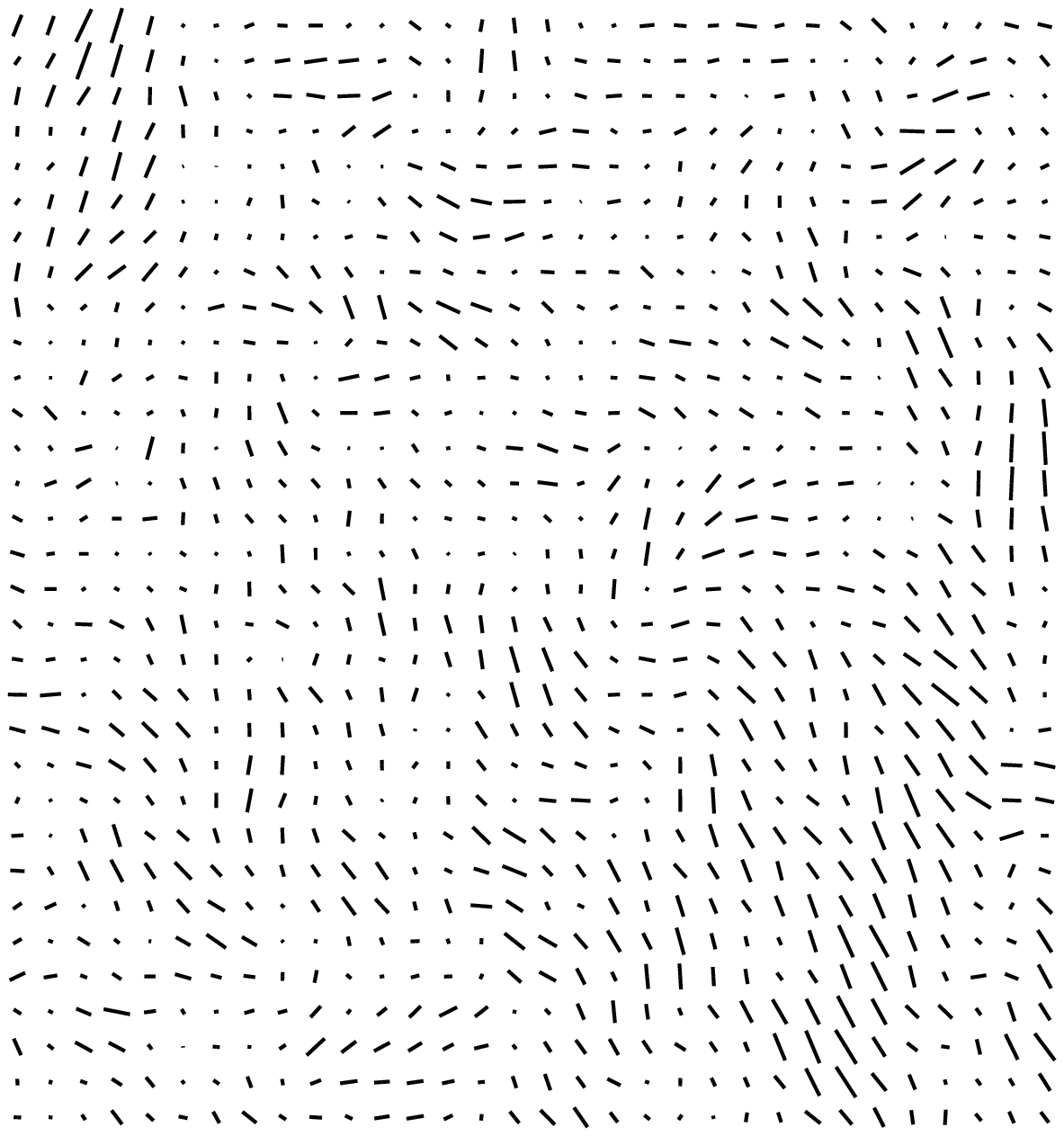}}
\caption{Example of real-space $E$/$B$/ambiguous decomposition.  The
original map (a) was decomposed into $E$ and $B$ components (b,c).
The $B$ component was then separated into ambiguous (d) and pure (e)
parts.  For comparison, the $B$ component of the true, original map
is shown in panel (f).  Because the amplitudes of the $B$ modes
was chosen to be lower than the $E$ modes by a factor 10, the polarizations
in 
panels (e,f) have been multiplied by 10.}
\label{fig:separation}
\end{figure}

Figure \ref{fig:separation} shows an example of this procedure.

This ``purification'' method may prove useful in conjunction with other
CMB analysis methods, such as the pseudo-$C_l$ method.
One proposed way of applying this method to polarization data
\cite{smithzal} is to apply the ``curl'' differential operator
${\bf D}_B$ to the original sky map before calculating pseudo-$C_l$'s.  Because
all $E$ modes (pure or otherwise) are mapped to zero by this operator,
this has the effect of projecting down onto the pure $B$ subspace.
Because this approach involves taking derivatives of noisy, pixelized data,
careful apodization of the weight functions used in calculating the 
pseudo-$C_l$'s is required.  The authors of this work have shown how
to solve this problem.  However, it is tempting to speculate
that the method would be easier to implement and more transparent
if differentiation of the data could be avoided from the beginning.
It may be possible to achieve this goal by performing the projection
described in this section instead of applying the operator ${\bf D}_B$.  
Both procedures have the effect of projecting down to the pure $B$ subspace,
but the procedure described here results in the actual pure $B$ map,
rather than a second derivative of it.  I wish to emphasize that this
proposal is at this stage largely speculation; this approach
requires further investigation.

\section{Discussion}

Separation into $E$ and $B$ components will be an essential step in the
analysis of future CMB polarization experiments, particularly those focused
on characterization of the weaker $B$ modes.  Fortunately, the $E$/$B$ separation problem does not pose major
data analysis challenges: methods for performing the separation in real
space are known, as are methods for calculating power spectrum estimates
on the original maps that take full account of the presence
of ambiguous modes (e.g., \cite{smithzal}).

When designing and optimizing
experiments, it is important to consider the degree to which $B$ mode
information is lost to ambiguous modes.
In general, the ambiguous modes primarily affect the largest observable
scales in any given experiment.  Unfortunately, these are also the scales
where the signal-to-noise is highest.  Because the ambiguous modes
are caused by conditions on the boundary of the map, the most efficient
survey designs are those with short boundaries (i.e., rounder is better).
Furthermore, aliasing of modes above the Nyquist frequency strongly
mixes the $E$ and $B$ components.  As a result, oversampling of the beam
is even more important in polarization experiments than in temperature
anisotropy measurements.

Although it is always possible to construct a basis of orthonormal modes
that implement the separation into $E,B$, and ambiguous components,
such a construction is cumbersome and computationally expensive in practice.
I have sketched a procedure for ``purifying'' the components of a data
set in real space without the construction of such a basis.  This 
approach may be useful when using estimators such as the pseudo-$C_l$'s,
although further research in this area is needed.

Finally, I would like to comment briefly on the tradeoffs between 
interferometric and single-dish imaging experiments.  
Interferometric visibilities are well-localized in Fourier space,
and hence they provide sharp power spectrum estimates \cite{parkng}.
Since, as we have seen above, $E$/$B$ mixing is closely connected with
resolution in Fourier space, we might expect interferometers to be
better than imaging experiments in this regard as well.  Further
research is needed to quantitatively assess this claim, however.

An individual pair of visibilities $(V_Q,V_U)$ from a single interferometer
baseline does contain separable information about $E$ and $B$ modes,
whereas no $E,B$ information can be extracted from a single pixel
of an imaging experiment.  For relatively short baselines, relatively
little $B$ information can be extracted from a single visibility
pair, however, because there is significant $E$/$B$ mixing.
In practice, $B$ information can be extracted from a single visibility
only if $\overline{s^2}<(C_B/C_E)\sim 10^{-2}$, where $\overline{s^2}$
is a measure of the spread in wavevector directions probed by the 
given baseline \cite{polsys}.  This inequality is achieved when
the antenna separation is at least $\sim 4$ times the antenna diameter.
Since CMB interferometers are traditionally close-packed, many
baselines are shorter than this.  $E$/$B$ separation is still
possible for these short baselines, but it requires dense sampling
of the visibility plane and/or mosaicking \cite{bunnwhite} 
to improve Fourier space
resolution.

\ack{This work was supported by National Science Foundation Awards 
AST-0098048 and AST-0507395.}

\bibliography{bunneb}

\end{document}